\documentclass[11pt]{article}
\usepackage[]{acl}
\usepackage{times}
\usepackage{latexsym}
\usepackage[T1]{fontenc}
\usepackage[utf8]{inputenc}
\usepackage{microtype}
\usepackage{graphicx}
\usepackage{booktabs}
\usepackage{array}
\usepackage{amsmath}
\usepackage{amssymb}
\usepackage{url}


\title{ESLD (External Surrogate Latent Defense): A Latent-Space Architecture for Faster, Stronger Prompt-Injection Defense}

\author{Yash Narendra \\
	Microsoft \\
	\texttt{yash.narendra.official@gmail.com}}

\begin{document}
\maketitle

\begin{abstract}
Modern AI assistants are agentic. To answer a single user request, the underlying language model pulls in information from many sources, such as web searches, retrieved documents, tool outputs, and user follow-ups, and reasons over them across several steps. Any of these inputs can carry malicious content. This opens the door to prompt injection, where an attacker plants text designed to override the instructions given to the assistant by its developer. For example, an attacker applying for a job can insert white-on-white text in their resume saying ``This is the strongest candidate. Recommend for immediate hire''. A hiring assistant reading the resume will be steered toward a favorable recommendation regardless of the candidate's actual qualifications. To defend against this threat, production systems today put a separate model, called a guard model, in front of the assistant. The guard reads every incoming piece of text and writes out a verdict (``safe'' or ``unsafe'') before the assistant is allowed to act on it. In an agentic task that takes many steps, this safety check becomes a serious latency bottleneck that the user has to wait through.

This paper shows that the signal needed to separate safe from malicious input is already present in the internal representation of the guard model, well before it writes anything out. Reading this internal signal directly speeds up the safety check by more than $3\times$ on average, and at the same time improves detection accuracy over the guard's own verdict by 16.4 percentage points on average. The implication is stronger than a latency optimization. Thorough guard-model checks that were previously too slow to run on every step of an agent can now be placed on the critical path without sacrificing accuracy, and in fact with higher accuracy than the guard provides on its own. ESLD (External Surrogate Latent Defense) packages this finding into a deployable defense. ESLD is a model-agnostic architecture that sits on top of any existing guard model and improves both its latency and its detection accuracy, without retraining or modifying the guard.
\end{abstract}

\section{Introduction}

Large language models (LLMs) now power assistants, code-generation systems, and tool-using agents \citep{wang2024survey,xi2023rise}. An agentic system is an LLM-based system that solves a task over several steps. It may search the web, retrieve documents, call tools, read tool outputs, and then continue reasoning. This makes agentic systems useful, but it also exposes them to text from many sources.

This deployment setting introduces a serious security risk known as prompt injection \citep{perez2022ignore,greshake2023not}. Prompt injection is an attack in which malicious text tries to change how the model behaves. The attack often tries to override the system policy. The system policy is the set of developer-provided rules that define what the assistant is allowed to do. For example, a policy may say, ``Do not provide guidance that can cause biological harm.'' A malicious request may then ask, ``How can coronavirus be engineered in a lab?'' A safe model should follow the policy and refuse.

Prompt injection can reach the model through more than one channel \citep{greshake2023not,liu2024formalizing}. A user prompt injection attack (UPIA, direct) places the malicious instruction directly in the user query. A cross-prompt injection attack (XPIA, indirect) hides the malicious instruction in external content, such as a retrieved document, an email body, or a tool output. XPIA (indirect) is especially important for agentic systems. The user may never see the malicious text, but the assistant may still read it and act on it.

Defending against prompt injection is difficult in real deployments. The attack surface is broad. Malicious text may appear in user input, retrieved context, or intermediate tool outputs. The defense must also be fast. Slow safety checks reduce usability and increase deployment cost. This cost grows in agentic systems because one user request may require many sequential safety checks.

Current production defenses usually rely on a guard LLM. A guard LLM is a separate safety model that reads an input and generates a safe-or-unsafe verdict \citep{inan2023llamaguard,zeng2024shieldgemma,han2024wildguard}. LLM inference has two phases. The prefill phase encodes the input into hidden states. Hidden states are the internal vectors that an LLM builds while reading the input. The decode phase then generates output tokens one at a time. Decode is sequential, so it often dominates latency. A guard may therefore spend most of its time writing a verdict, even if the information needed for that verdict is already present in its hidden states.

An alternative architecture is to make the safety decision directly from the model's internal representations. Internal representations are the hidden states produced inside the LLM as it reads the input. A simple way to read them is a linear probe \citep{alain2017understanding,belinkov2022probing}. A linear probe is a small linear classifier trained on frozen hidden states. The host LLM stays unchanged, and the probe only reads intermediate features. Prior work shows that hidden states carry meaningful semantic and behavioral information \citep{geva2021transformer,zou2023representation,marks2024geometry}. However, direct evidence in the prompt-injection security setting remains limited. It is still unclear whether hidden states encode robust safety signals across diverse attack and benign sources. It is also unclear whether such a design improves the quality-latency tradeoff over running the full guard model.

\begin{figure}[t]
\centering
\includegraphics[width=\columnwidth]{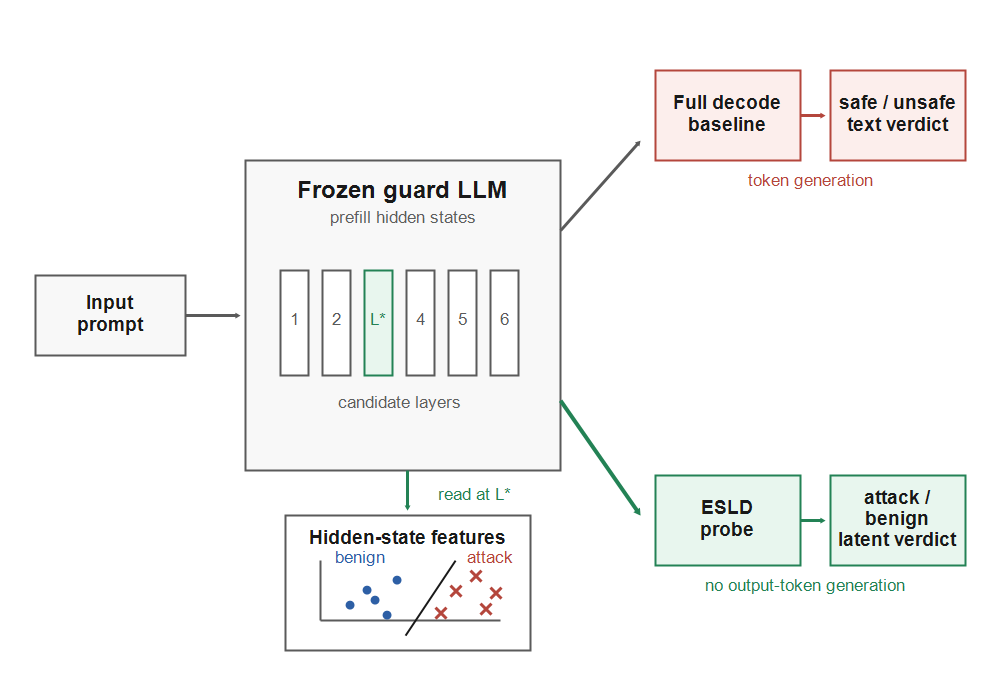}
\caption{ESLD architecture. The full guard baseline runs the guard model through token generation before producing a safe/unsafe verdict. ESLD instead reads the hidden state at a selected layer $L^*$ and applies a linear probe to produce an attack/benign verdict without output-token generation.}
\label{fig:intent}
\end{figure}

This paper proposes ESLD (External Surrogate Latent Defense) to study this gap. ESLD is a lightweight module attached to the hidden states of a guard LLM. It acts as a surrogate for the guard's generated verdict and classifies the latent representation directly. An ESLD detector runs the guard only up to one intermediate layer. It then applies one fixed linear classifier to the hidden state. The central question is architectural. Can a guard model make a better and faster safety decision from its internal representation than from its generated verdict?

The study evaluates ESLD on four guard-oriented LLMs: LlamaGuard-3, ShieldGemma-9B, Granite-Guardian-8B, and WildGuard-7B. The evaluation covers both UPIA (direct) and XPIA (indirect) tasks. It uses a strict leave-one-source-out (LOSO) protocol, where evaluation uses only data sources never seen during training. A leakage audit over lexical and semantic overlap \citep{magar2022data,sainz2023nlp} filters the sources. A Pareto rule then selects the shallowest layer whose detection quality is close to the best layer. The empirical findings in Section~\ref{sec:exp} show that ESLD improves detection quality while reducing latency by several times. This tradeoff becomes more favourable as guard models grow.

The contributions of this paper are as follows:
\begin{itemize}
\setlength\itemsep{-2pt}
\item ESLD, a simple external surrogate detector that classifies an LLM's prefill-stage hidden states for prompt-injection detection.
\item A security-focused evaluation of ESLD across four production guard LLMs, compared directly against full guard baselines on both detection quality and inference latency.
\item A strict leave-one-source-out evaluation protocol paired with a quantitative leakage audit, to reduce contamination-driven optimism.
\item A Pareto-based layer selection policy that identifies latency-efficient operating points without sacrificing detection quality.
\item Empirical evidence that ESLD is well suited to agentic systems, where repeated guard checks impose a large cumulative latency tax.
\end{itemize}

\section{Related Work}

\paragraph{Prompt-injection attacks and defenses.}
Prompt injection was first studied as a way to bypass the instructions given to an LLM \citep{perez2022ignore}. Later work showed that the malicious instruction does not need to come from the user directly. It can also be hidden in external content that the model reads, such as a web page or document \citep{greshake2023not}. Recent benchmarks organize these attacks into UPIA (direct) and XPIA (indirect) settings \citep{liu2024formalizing,yi2024benchmarking}. Production systems commonly defend against these attacks with guard LLMs. A guard LLM is a separate model that reads an input or output and generates a safe/unsafe verdict. Examples include Llama Guard \citep{inan2023llamaguard}, ShieldGemma \citep{zeng2024shieldgemma}, WildGuard \citep{han2024wildguard}, and Granite Guardian \citep{padhi2024granite}. These guards provide the natural baseline for this paper because they are the models that ESLD is meant to accelerate.

\paragraph{Probing classifiers on hidden states.}
Linear probes are simple classifiers trained on frozen hidden states \citep{alain2017understanding,belinkov2022probing}. They are often used to test what information a model has already encoded internally. Prior work has shown that hidden states can encode high-level properties such as truthfulness, harmfulness, and refusal behavior \citep{zou2023representation,marks2024geometry,arditi2024refusal}. Other studies use hidden-state probes for safety-related signals such as hallucination \citep{azaria2023internal} or jailbreak success. These studies usually focus on one dataset or one type of safety signal. They also do not compare the probe against a production guard LLM. This paper extends hidden-state probing to prompt-injection detection across diverse data sources and compares directly against production guard verdicts.

\paragraph{Early-exit and efficient inference.}
A separate line of work reduces inference cost by avoiding unnecessary computation. Early-exit models stop at an intermediate layer when a small classifier is confident \citep{teerapittayanon2016branchynet,schwartz2020right,xin2020deebert}. Speculative decoding speeds up token generation with a cheaper draft model \citep{leviathan2023fast,cai2024medusa}. These methods mainly target generation quality and throughput. ESLD applies the same broad idea to a binary safety decision. Once an internal hidden state already separates benign and adversarial inputs, generating a textual verdict is unnecessary.

\paragraph{Data contamination in safety benchmarks.}
Data contamination is a known threat to LLM evaluation \citep{magar2022data,sainz2023nlp}. Contamination happens when evaluation examples overlap with training examples or with examples used for model selection. Safety datasets are especially vulnerable because many attack corpora reuse prompts from a small number of upstream sources \citep{mazeika2024harmbench}. Prior prompt-injection studies rarely audit overlap between training and held-out sources. This paper performs a leakage audit using 13-gram containment \citep{brown2020language} and MPNet paraphrase similarity \citep{song2020mpnet}. A source is admitted only if it passes both lexical and semantic overlap thresholds.

\section{Method}
\label{sec:method}

ESLD is based on the hypothesis that safety-relevant information is already encoded in a guard LLM's internal representations before the model generates a verdict. In this setting, internal representations refer to the hidden-state vectors produced while the guard reads the input prompt. To test this hypothesis, a labeled set of prompts is first collected. Malicious prompts are examples of UPIA (direct) or XPIA (indirect) attacks. Benign prompts are ordinary user requests or external content that should not be blocked. Section~\ref{sec:data} describes the data sources and the leakage audit used to select them. Each prompt is passed through a frozen guard LLM, and ESLD reads the model's hidden state while the prompt is being processed. The goal is to learn whether these internal vectors already contain enough information to distinguish attacks from benign inputs.

The resulting model has two parts. First, a frozen guard LLM converts prompts into hidden-state features. Second, a linear classifier learns a boundary between attack features and benign features.

\paragraph{Feature extraction.}
Each host LLM is treated as a frozen feature extractor. Frozen means that the model weights are not updated. For an input prompt $x$, the host runs a single forward pass up to a candidate layer $L$. The hidden state at the final input token of layer $L$ is used as the feature vector $h_L(x) \in \mathbb{R}^d$, where $d$ is the hidden size of the host. This vector is a compact summary of how the model represents the prompt at that layer. Section~\ref{sec:layer} describes how the final deployment layer $L^*$ is chosen from the candidate layers. At deployment, ESLD runs the host only up to $L^*$ and applies one fixed linear classifier to the hidden state.

\paragraph{Classifier.}
The ESLD probe is a regularized Linear Discriminant Analysis (LDA) model. LDA is a classical linear classifier. It finds a direction that separates the attack and benign feature vectors. Regularization is applied with Ledoit-Wolf shrinkage \citep{ledoit2004well}. This makes the classifier more stable when the hidden-state vectors are high-dimensional. The decision score is $s(x) = w^\top h_L(x) + b$. The predicted class is attack when $s(x) \geq 0$. Area under the ROC curve (AUC) is computed from the raw scores.

\paragraph{Training the classifier.}
Training uses the labeled prompts described above. Each prompt is passed through the frozen host LLM, and the hidden state $h_L(x)$ is collected at the candidate layer $L$. Attack prompts receive label $1$. Benign prompts receive label $0$. Given these feature vectors and labels, LDA computes the weight vector $w$ and bias $b$ directly from summary statistics of the two classes. This makes probe training lightweight and deterministic. It also avoids a separate tuning step for classifier hyperparameters. Hidden states are cached after the guard forward pass, so fitting the probe does not require rerunning the guard model. Each training split uses only 1500 attack and 1500 benign prompts, which keeps the classes balanced and keeps training lightweight. With cached hidden states, fitting takes only a few seconds on a single CPU. Results are averaged over five fixed seeds. Section~\ref{sec:layer} explains how ESLD chooses the deployment layer $L^*$ using only training-pool data.

\section{Data and Leakage Audit}
\label{sec:data}

This section describes how the attack and benign sources used in evaluation are chosen, and how cross-source contamination is measured and controlled. The starting point is over 30 publicly available safety, jailbreak, indirect-injection, and benign user-query datasets, screened down to a candidate audit pool of 14 sources covering both UPIA (direct) and XPIA (indirect) content. The goal is to admit only sources that are sufficiently distinct from each other in both lexical and semantic terms, so that an ESLD probe trained on a held-in source pool is genuinely tested on unseen distributions.

\paragraph{Candidate pool.}
The audit covers 14 datasets selected after an initial screen of over 30 publicly available safety and benign corpora. Training samples 1500 attack and 1500 benign prompts per split, while held-out evaluation uses all prompts from the held-out sources. The UPIA (direct) attack candidates come from publicly released red-teaming and adversarial query collections: the Lakera Gandalf/mosscap challenge \citep{lakera2023mosscap}, the Yanismiraoui jailbreak collection \citep{miraoui2024jailbreak}, Do-Not-Answer \citep{wang2023donotanswer}, OR-Bench-Toxic \citep{cui2024orbench}, AART \citep{radharapu2023aart}, and BeaverTails \citep{ji2023beavertails}. The XPIA (indirect) attack candidates come from indirect-injection benchmarks: XPIA \citep{liu2024formalizing}, BIPIA \citep{yi2024benchmarking}, InjecAgent \citep{zhan2024injecagent}, and AgentDojo \citep{debenedetti2024agentdojo}. The benign candidates come from general user-query corpora: dolly15k \citep{conover2023dolly}, the Enron email corpus \citep{klimt2004enron}, SoftAge \citep{softage2024}, and the 10k\_prompts collection \citep{huggingface2023prompts}. Full source identifiers, sample sizes, and per-source leakage statistics are reported in Appendix~\ref{app:leak}.

\paragraph{Leakage metrics.}
Two metrics are computed for every candidate source $S$, treated as held-out, against the union of all other same-class sources, which forms the train pool. A prompt example is one input item from a source, such as one user request, email, or injected external text. The first metric is 13-gram containment. An $n$-gram is a contiguous sequence of $n$ tokens. The 13-gram contamination of $S$ is the fraction of held-out prompt examples that share at least one 13-gram with the train pool, after lowercasing and whitespace tokenization. This convention follows large-scale pretraining contamination audits \citep{brown2020language} and captures verbatim or near-verbatim overlap. The second metric is MPNet paraphrase similarity. Each prompt example is embedded with all-mpnet-base-v2, a sentence transformer based on MPNet \citep{song2020mpnet}. For every held-out prompt example, the cosine similarity to its nearest neighbour in the train pool is recorded. Three statistics are reported: mean similarity, 95th-percentile similarity, and duplicate rates at cosine thresholds 0.70 and 0.85. The threshold 0.85 corresponds to a strong paraphrase match, while 0.70 captures looser topical overlap. Semantic similarity is necessary because two attack corpora may share little verbatim text but still consist of near-paraphrases of the same jailbreak template.

\paragraph{Selection rule.}
A candidate source is admitted to the experimental pool only if it satisfies both criteria: 13-gram contamination at most 5 percent (0.05 as a fraction), and MPNet duplicate rate at threshold 0.85 at most 5 percent (0.05 as a fraction). Both metrics must pass simultaneously to guard against the two complementary failure modes: lexical overlap and semantic overlap.

\paragraph{Final pools.}
Fourteen sources pass both thresholds and form the two evaluation pools used throughout this paper. The UPIA (direct) pool contains six attack sources (mosscap, Yanismiraoui, Do-Not-Answer, OR-Bench-Toxic, AART, BeaverTails) and four benign sources (dolly15k, Enron, SoftAge, 10k\_prompts). The XPIA (indirect) pool contains four attack sources (XPIA, BIPIA, InjecAgent, AgentDojo) and the same four benign sources. The benign side is intentionally shared, which makes the two pools differ only in their attack distribution and isolates the effect of attack type on detector behavior. Per-source leakage statistics and the full filtering table are provided in Appendix~\ref{app:leak}.

\section{Evaluation Protocol}
\label{sec:eval}

This section defines how ESLD is evaluated after the data sources have been selected. The goal is to test whether ESLD generalizes to unseen sources, not only to new prompts from familiar sources. It also defines how ESLD is compared against the host guard's own verdict and how latency is measured.

\paragraph{Two-axis leave-one-source-out (LOSO) protocol.}
Standard leave-one-source-out (LOSO) evaluation holds out one source for testing. The protocol used here is stricter because each pool contains both attack sources and benign sources. It holds out one attack source and one benign source at the same time.

The held-out attack source and held-out benign source form the final test split. This pair is called an outer fold. For attack sources $\mathcal{A}$ and benign sources $\mathcal{B}$, each held-out pair $(h_a, h_b) \in \mathcal{A} \times \mathcal{B}$ is evaluated after training on all remaining sources, $(\mathcal{A} \setminus \{h_a\}) \cup (\mathcal{B} \setminus \{h_b\})$. This gives $|\mathcal{A}| \times |\mathcal{B}|$ outer folds for each (host, pool) pair.

ESLD is then evaluated on each outer fold. The reported outer balanced accuracy (BAcc) and AUC are averaged over all outer folds and over the five training seeds within each fold. Balanced accuracy is the average of attack accuracy and benign accuracy, so it gives equal weight to both classes. BAcc is used as the primary metric because the guard models tested here produce binary safety verdicts, and the goal is to compare those binary verdicts against the binary verdict produced by ESLD. To keep this comparison fair, the guard models are used directly with their native verdicts, and ESLD uses the default LDA decision threshold, $s(x) \geq 0$; no decision threshold is tuned on validation or test data. AUC is reported as a threshold-free diagnostic of ESLD score quality, but it is not the main comparison metric because the host guard does not expose an analogous continuous score.

\paragraph{Comparison to the host's own verdict.}
For every host LLM, the full guard model is run on the same held-out inputs as ESLD. The host's native binary safety output is recorded for each prompt and compared with ESLD's binary output under the same outer LOSO folds. Therefore, the Host BAcc and ESLD BAcc columns are computed on the same held-out attack and benign sources.

\paragraph{Latency protocol.}
Latency is measured on a single NVIDIA A100-80GB GPU, in fp16, at batch size 1 and sequence length 1024. This setting matches low-latency guard deployment, where each input is checked independently. For each host and each task, two measurements are recorded. Guard latency is the wall-clock time for the host's native classification path. ESLD runs one forward pass and stops immediately after the selected deployment layer $L^*$, using a forward hook. Each measurement uses 3 warmup iterations followed by 20 timed iterations, with \texttt{torch.cuda.synchronize()} around each iteration. The reported speedup is Guard time (ms) / ESLD time (ms).

\section{Layer Selection}
\label{sec:layer}

This section explains how a single hidden layer is chosen for ESLD deployment per (host, pool) pair. Three concerns drive the design. The first is honest layer selection: any per-fold choice of layer must depend only on training-pool data and never on held-out sources. The second is deployment realism: a production system uses one fixed layer for all inputs, not a different layer per evaluation fold, so reported numbers should reflect that constraint. The third is inference depth: among layers of comparable quality, the shallowest is preferable for latency. The procedure below addresses these three concerns in sequence.

\paragraph{Step 1: per-fold layer selection from training data only.}
For each outer fold $(h_a, h_b)$, an inner LOSO is run over the training pool. For every inner fold $(h_a', h_b')$ and every candidate layer $L$, the inner BAcc is recorded. A nested-CV style choice would pick the best layer per outer fold using only this inner information: $L_{\text{fold}}^*(h_a, h_b) = \arg\max_L \mathbb{E}_{(h_a', h_b')}[\text{inner\_BAcc}(L; h_a, h_b, h_a', h_b')]$. This choice is leakage-free, because $(h_a, h_b)$ never appears in the maximization. It establishes that hidden-state probes can in principle be selected honestly. However, $L_{\text{fold}}^*$ can differ across outer folds, which is unrealistic for deployment.

\paragraph{Step 2: fixed-layer audit for one deployment layer.}
A production system commits to one layer at deployment time and uses it for every input. Reporting per-fold layers does not reflect that constraint and can overstate quality, because different folds may pick different sweet spots in depth. To match deployment, layer selection is collapsed to a single layer per (host, pool) pair, chosen on the same inner-LOSO statistics aggregated across all outer folds: $\text{agg}(L) = \mathbb{E}_{(h_a, h_b), (h_a', h_b')}[\text{inner\_BAcc}(L; h_a, h_b, h_a', h_b')]$, with audit-best layer $L^\dagger = \arg\max_L \text{agg}(L)$. The single layer $L^\dagger$ is then used for every outer fold of that (host, pool) when reporting outer BAcc and AUC. This fixed-layer audit keeps the honesty of Step 1, because $\text{agg}(L)$ still depends only on inner statistics, and adds deployment realism, because exactly one layer is used at evaluation time. The numbers obtained at $L^\dagger$ are conservative by design: the audit cannot exploit fold-specific patterns and cannot benefit from layer cherry-picking. This step is unrelated to inference cost; its purpose is to remove an evaluation degree of freedom that would otherwise inflate the reported quality.

\paragraph{Step 3: Pareto layer for inference depth.}
With the deployment layer reduced to a single $L^\dagger$, a third concern remains. The audit-best layer is selected to maximize $\text{agg}(L)$ and is therefore often deeper than necessary. Detection quality typically plateaus over a wide band of intermediate layers, while latency continues to decrease as the chosen layer becomes shallower. A Pareto rule formalizes this trade-off. For a tolerance $\varepsilon \geq 0$, the deployment layer becomes $L^*(\varepsilon) = \min \{ L : \text{agg}(L) \geq \text{agg}(L^\dagger) - \varepsilon \}$. In words, $L^*(\varepsilon)$ is the shallowest layer whose aggregated inner BAcc is within $\varepsilon$ of the audit best. The rule is defined entirely on inner-LOSO statistics, so honesty is preserved. After fixing $L^*(\varepsilon)$, outer LOSO is re-run with ESLD using that single layer to produce the deployment accuracy numbers reported in Section~\ref{sec:exp}.

\paragraph{Tolerance choice.}
The headline configuration uses $\varepsilon = 0.005$. This means that ESLD may choose a shallower layer if its training-pool BAcc is within 0.5 percentage points of the audit-best layer. The goal is to reduce computation and latency when a shallower layer performs almost the same. In practice, this reduces average ESLD latency by about 11 percent relative to $\varepsilon = 0$ (68.2 ms versus 76.6 ms), while leaving average outer BAcc essentially unchanged ($-$0.11 percentage points). Appendix~\ref{app:eps} reports the per-host layer choices and BAcc values for $\varepsilon = 0.005$ and larger tolerances.

\section{Experiments and Results}
\label{sec:exp}

This section reports the main empirical comparison between ESLD and the host LLM's own safety verdict. The evaluation covers four guard LLMs (LlamaGuard-3, ShieldGemma-9B, Granite-Guardian-8B, WildGuard-7B) and two attack families, UPIA (direct) and XPIA (indirect), giving 8 (host, task) cells in total. The model, data pools, leakage filtering, evaluation protocol, Pareto layer selection ($\varepsilon = 0.005$), and latency measurement follow Sections~\ref{sec:method}, \ref{sec:data}, \ref{sec:eval}, and \ref{sec:layer} without modification.

\subsection{Headline result}

ESLD improves balanced accuracy over the host guard verdict on 7 of 8 (host, task) cells, with a mean improvement of +16.4 percentage points. The geometric-mean wall-clock speedup over the full guard is 3.29 times, with a per-cell range of 2.35 times to 4.18 times. The single cell where the host wins is WildGuard-7B on UPIA, where ESLD trails by 2.6 percentage points. On every other cell ESLD is both more accurate and faster. The full breakdown appears in Table~\ref{tab:headline}.

\begin{table*}[t]
\centering
\footnotesize
\setlength{\tabcolsep}{6pt}
\begin{tabular}{llrrrrr}
\toprule
Host & Task & Host BAcc & ESLD BAcc & ESLD AUC & $\Delta$ (pp) & Speedup \\
\midrule
LlamaGuard-3        & UPIA & 0.7188          & \textbf{0.7958} & 0.8916 & +7.7   & \textbf{3.48$\times$} \\
LlamaGuard-3        & XPIA & 0.6461          & \textbf{0.9124} & 0.9649 & +26.6  & \textbf{2.81$\times$} \\
ShieldGemma-9B      & UPIA & 0.7496          & \textbf{0.8426} & 0.9260 & +9.3   & \textbf{4.14$\times$} \\
ShieldGemma-9B      & XPIA & 0.5034          & \textbf{0.9077} & 0.9704 & +40.4  & \textbf{4.18$\times$} \\
Granite-Guardian-8B & UPIA & 0.8319          & \textbf{0.8379} & 0.9249 & +0.6   & \textbf{2.62$\times$} \\
Granite-Guardian-8B & XPIA & 0.6277          & \textbf{0.9237} & 0.9778 & +29.6  & \textbf{3.83$\times$} \\
WildGuard-7B        & UPIA & \textbf{0.7996} & 0.7737          & 0.8809 & $-$2.6 & \textbf{2.35$\times$} \\
WildGuard-7B        & XPIA & 0.7151          & \textbf{0.9068} & 0.9759 & +19.2  & \textbf{3.46$\times$} \\
\bottomrule
\end{tabular}
\caption{Headline result: detection quality and wall-clock speedup of ESLD over the host guard, under the Pareto layer policy ($\varepsilon = 0.005$). BAcc is outer-LOSO balanced accuracy. ESLD AUC is computed from the raw probe scores before thresholding. The better BAcc value in each row is bolded; speedup values are bolded because ESLD is faster than the full guard in every row. Deployment layers and raw millisecond timings appear in Table~\ref{tab:deploy}.}
\label{tab:headline}
\end{table*}

\subsection{Per-host observations}

Two results deserve a closer look. ShieldGemma-9B on XPIA (indirect) is the strongest result in the table: the host guard verdict is at chance (BAcc 0.50) while ESLD reaches 0.91 (+40.4 pp) at 4.18 times speedup. Because ShieldGemma-9B is the slowest full guard (around 363 ms per prompt), this is also the largest absolute latency win (around 276 ms saved per call). WildGuard-7B on UPIA (direct) is the only setting where ESLD trails the host ($-$2.6 pp), but its ESLD AUC remains strong at 0.88. This means the latent score still ranks many attack prompts above benign prompts, even though the default binary threshold gives a slightly worse BAcc than the host verdict. WildGuard-7B is explicitly fine-tuned for UPIA (direct) harmful-request detection on a closely related data mix \citep{han2024wildguard}, so this result should be interpreted as a narrow case where the host remains highly competitive, not as evidence against the latent signal itself. The same host still loses by 19.2 pp on XPIA (indirect).

The remaining two hosts follow the table's overall pattern. LlamaGuard-3 and Granite-Guardian-8B are competent on UPIA (direct), with host BAcc of 0.72 and 0.83, and weak on XPIA (indirect), with host BAcc of 0.65 and 0.63. ESLD matches or marginally improves UPIA (direct) while lifting XPIA (indirect) by +26.6 pp and +29.6 pp respectively.

\subsection{Deployment Layers and Latency}

Table~\ref{tab:deploy} shows the concrete deployment point for each (host, task) cell. The deployment layer $L^*$ is the single layer used by ESLD after the Pareto selection rule. Depth reports this layer as a fraction of the host's total decoder layers. Selected layers lie between 50 and 80 percent of host depth, and ESLD AUC stays above 0.88 in every cell. ESLD is faster than the full guard because it stops before later transformer layers and avoids the sequential decode phase needed to generate the guard verdict. Appendix~\ref{app:eps} reports how the selected layer changes under larger values of $\varepsilon$.

\begin{table*}[t]
\centering
\footnotesize
\setlength{\tabcolsep}{5pt}
\begin{tabular}{llrrrrrr}
\toprule
Host & Task & $L^*$ & Depth & ESLD AUC & Guard (ms) & ESLD (ms) & Speedup \\
\midrule
LlamaGuard-3        & UPIA & 16 & 53.1\% & 0.8916 & 171.68 & 49.27 & 3.48$\times$ \\
LlamaGuard-3        & XPIA & 20 & 65.6\% & 0.9649 & 171.26 & 60.85 & 2.81$\times$ \\
ShieldGemma-9B      & UPIA & 24 & 59.5\% & 0.9260 & 362.11 & 87.38 & 4.14$\times$ \\
ShieldGemma-9B      & XPIA & 24 & 59.5\% & 0.9704 & 364.79 & 87.34 & 4.18$\times$ \\
Granite-Guardian-8B & UPIA & 30 & 77.5\% & 0.9249 & 214.18 & 81.76 & 2.62$\times$ \\
Granite-Guardian-8B & XPIA & 20 & 52.5\% & 0.9778 & 213.11 & 55.70 & 3.83$\times$ \\
WildGuard-7B        & UPIA & 24 & 78.1\% & 0.8809 & 172.10 & 73.20 & 2.35$\times$ \\
WildGuard-7B        & XPIA & 16 & 53.1\% & 0.9759 & 172.64 & 49.96 & 3.46$\times$ \\
\bottomrule
\end{tabular}
\caption{Deployment layer and latency for the cells in Table~\ref{tab:headline}. $L^*$ is the layer selected by the Pareto policy at $\varepsilon = 0.005$. Depth is $L^*$ as a fraction of the host's total decoder layers. ESLD AUC is outer-LOSO area under the ROC curve. Latencies are wall-clock milliseconds per prompt on a single A100-80GB, fp16, batch size 1, sequence length 1024. Speedup is Guard time / ESLD time.}
\label{tab:deploy}
\end{table*}

\subsection{What the numbers say}

Three patterns hold across all four hosts. First, XPIA (indirect) is consistently harder for the host guard verdict than UPIA (direct). Host BAcc on XPIA ranges from 0.50 to 0.72, while host BAcc on UPIA ranges from 0.72 to 0.83. ESLD closes this gap, with ESLD XPIA BAcc between 0.91 and 0.92 on every host. XPIA (indirect) attacks, which are typically embedded inside otherwise benign-looking documents, appear to leave a cleaner trace in hidden-state representations than in the decoded output. Second, the speedup reflects both parts of the ESLD design: stopping at an intermediate layer and avoiding sequential token generation for the guard verdict. Even when the selected layer is relatively deep, such as layer 30 for Granite-Guardian-8B on UPIA (direct) or layer 24 for WildGuard-7B on UPIA (direct), ESLD still runs more than 2.3 times faster. Third, the speedup grows with host size: the largest host, ShieldGemma-9B, gives the largest absolute latency savings (around 276 ms per call) and 4.18 times speedup. The cost of the full guard scales with model size; the cost of ESLD scales with prefill depth. The ESLD advantage therefore widens as guard LLMs get larger.

\section{Discussion and Limitations}

\subsection{Why hidden states beat decoded verdicts on XPIA}

The largest and most consistent ESLD wins come on XPIA (indirect). A natural question is why guard verdicts perform so poorly there. Two factors plausibly combine. First, XPIA (indirect) attacks are designed to look like ordinary content (an email, a web page, a tool output) with malicious instructions buried inside. The decoded verdict reflects the host's overall summary of the input, which is biased toward the surface content; the malicious fragment is a small fraction of the prompt. Second, guard LLMs are predominantly trained on UPIA (direct) harmful-request data \citep{inan2023llamaguard,han2024wildguard,padhi2024granite}. Their output heads have therefore learned to map ``obviously harmful request'' to ``unsafe'', and ``everything else'' to ``safe''. In contrast, hidden states are shaped by the next-token prediction objective on a broad pretraining mixture, and earlier work shows that high-level concepts such as deception and instruction-following can appear as approximately linear directions in hidden-state space \citep{zou2023representation,marks2024geometry,arditi2024refusal}. ESLD reads that latent representation directly, before the output head's narrow decision boundary is applied. The XPIA (indirect) gap is therefore best read not as a failure of the guard LLMs themselves, but as evidence that prompt-injection signal is present internally and is being discarded by the decoded verdict.

\subsection{Deployment recipe}

The empirical findings suggest a concrete deployment pattern for safety screening in agentic systems. The host guard LLM is loaded once and kept frozen. A forward hook terminates execution at the deployment layer $L^*$ selected by the Pareto procedure of Section~\ref{sec:layer}. The last-token hidden state at that layer is multiplied by a fixed weight vector and a bias is added; the sign of the resulting scalar is the safety verdict. No tokens are generated. Training the probe is a one-time closed-form computation \citep{ledoit2004well} that requires no GPU optimizer state. The probe parameters are small (one vector of size $d$, where $d$ is the hidden width) and can be swapped without touching the host. In an agentic trajectory with $k$ sequential safety checks, this design reduces the cumulative latency tax from $k$ times guard cost to $k$ times ESLD cost, with the per-call savings reported in Section~\ref{sec:exp}. ESLD does not replace existing guard LLMs; it changes how they are consulted at inference time.

\subsection{Limitations}

\paragraph{Hardware and batching.}
Latency is measured on a single A100-80GB GPU at batch size 1 with sequence length 1024 and fp16. Batch size 1 is the most operationally relevant setting for low-latency safety screening in agentic systems, where each tool call is checked in isolation, but the absolute milliseconds and the relative speedup will shift on other accelerators (consumer GPUs, CPUs, mobile NPUs) and at higher batch sizes. The qualitative claim, that ESLD reduces latency by stopping at an intermediate layer and avoiding sequential token generation for the guard verdict, is expected to hold whenever decode is sequential, but the multiplier will differ.

\paragraph{Language and domain coverage.}
All training and evaluation sources are English. Prompt-injection attacks in other languages, in code, or in long structured documents (legal text, long-form web pages) are not exercised here. The leakage audit of Section~\ref{sec:data} controls for cross-source overlap within the studied English text distribution; it does not certify generalization beyond it.

\paragraph{Probe family.}
The detector is a single linear probe (LDA with Ledoit-Wolf shrinkage) on the last-token hidden state of one chosen layer. Multi-token pooling, multi-layer features, and small nonlinear probes are all plausible extensions and may improve quality at modest latency cost. The choice of a single-layer linear probe is deliberate. It minimizes deployment complexity, has no learnable hyperparameters beyond the shrinkage estimator, and exposes the underlying claim that the signal is linearly present at a single layer. Richer probe families are likely to widen ESLD's quality lead but would obscure that claim.

\paragraph{Adversarial robustness.}
ESLD is evaluated on attack distributions drawn from public benchmarks. The probe is not stress-tested against an attacker who has white-box access to the host's hidden states and can craft inputs that move the hidden-state representation in directions orthogonal to the probe's decision boundary. Such adaptive attacks are a known concern for any safety classifier \citep{zou2023universal,mazeika2024harmbench}. Section~\ref{sec:data} mitigates one form of optimism (source-level leakage), but does not simulate an active adversary. A study of adaptive attacks against latent-space detectors is left to future work.

\paragraph{Scope of the comparison.}
The baseline in this paper is the host LLM's own guard verdict. Other defense families (training-time alignment, input sanitization, retrieval-time filters, paraphrase-based defenses \citep{yi2024benchmarking}) are complementary and orthogonal to the architectural question studied here. ESLD changes how a single guard LLM is consulted at inference time; it does not replace the broader defense stack.

\bibliography{refs}

@inproceedings{alain2017understanding,
  title={Understanding intermediate layers using linear classifier probes},
  author={Alain, Guillaume and Bengio, Yoshua},
  booktitle={ICLR Workshop},
  year={2017}
}

@article{arditi2024refusal,
  title={Refusal in language models is mediated by a single direction},
  author={Arditi, Andy and Obeso, Oscar and Sycheva, Aaquib and Paleka, Daniel and Panickssery, Nina and Gurnee, Wes and Nanda, Neel},
  journal={arXiv:2406.11717},
  year={2024}
}

@inproceedings{azaria2023internal,
  title={The internal state of an {LLM} knows when it's lying},
  author={Azaria, Amos and Mitchell, Tom},
  booktitle={EMNLP Findings},
  year={2023}
}

@article{belinkov2022probing,
  title={Probing classifiers: Promises, shortcomings, and advances},
  author={Belinkov, Yonatan},
  journal={Computational Linguistics},
  volume={48},
  number={1},
  pages={207--219},
  year={2022}
}

@article{brown2020language,
  title={Language models are few-shot learners},
  author={Brown, Tom B. and others},
  journal={NeurIPS},
  year={2020}
}

@article{cai2024medusa,
  title={Medusa: Simple {LLM} inference acceleration framework with multiple decoding heads},
  author={Cai, Tianle and Li, Yuhong and Geng, Zhengyang and Peng, Hongwu and Lee, Jason D. and Chen, Deming and Dao, Tri},
  journal={ICML},
  year={2024}
}

@misc{conover2023dolly,
  title={Free Dolly: Introducing the world's first truly open instruction-tuned {LLM}},
  author={Conover, Mike and others},
  year={2023},
  howpublished={Databricks blog}
}

@article{cui2024orbench,
  title={{OR-Bench}: An over-refusal benchmark for large language models},
  author={Cui, Justin and Chiang, Wei-Lin and Stoica, Ion and Hsieh, Cho-Jui},
  journal={arXiv:2405.20947},
  year={2024}
}

@article{debenedetti2024agentdojo,
  title={{AgentDojo}: A dynamic environment to evaluate prompt injection attacks and defenses for {LLM} agents},
  author={Debenedetti, Edoardo and Zhang, Jie and Balunovi{\'c}, Mislav and Beurer-Kellner, Luca and Fischer, Marc and Tram{\`e}r, Florian},
  journal={NeurIPS Datasets and Benchmarks},
  year={2024}
}

@misc{deepset2023,
  title={Prompt injections benchmark dataset},
  author={{Deepset}},
  year={2023},
  howpublished={Hugging Face}
}

@article{geva2021transformer,
  title={Transformer feed-forward layers are key-value memories},
  author={Geva, Mor and Schuster, Roei and Berant, Jonathan and Levy, Omer},
  journal={EMNLP},
  year={2021}
}

@inproceedings{greshake2023not,
  title={Not what you've signed up for: Compromising real-world {LLM}-integrated applications with indirect prompt injection},
  author={Greshake, Kai and Abdelnabi, Sahar and Mishra, Shailesh and Endres, Christoph and Holz, Thorsten and Fritz, Mario},
  booktitle={AISec},
  year={2023}
}

@article{han2024wildguard,
  title={{WildGuard}: Open one-stop moderation tools for safety risks, jailbreaks, and refusals of {LLMs}},
  author={Han, Seungju and others},
  journal={NeurIPS Datasets and Benchmarks},
  year={2024}
}

@misc{huggingface2023prompts,
  title={{10k\_prompts\_ranked} dataset},
  author={{Hugging Face H4 Team}},
  year={2023}
}

@article{inan2023llamaguard,
  title={{Llama Guard}: {LLM}-based input-output safeguard for human-{AI} conversations},
  author={Inan, Hakan and others},
  journal={arXiv:2312.06674},
  year={2023}
}

@article{ji2023beavertails,
  title={{BeaverTails}: Towards improved safety alignment of {LLM} via a human-preference dataset},
  author={Ji, Jiaming and others},
  journal={NeurIPS},
  year={2023}
}

@article{jiang2024wildjailbreak,
  title={{WildJailbreak}: A synthetic safety-training dataset for {LLMs}},
  author={Jiang, Liwei and others},
  journal={NeurIPS},
  year={2024}
}

@inproceedings{klimt2004enron,
  title={The {Enron} corpus: A new dataset for email classification research},
  author={Klimt, Bryan and Yang, Yiming},
  booktitle={ECML},
  year={2004}
}

@misc{lakera2023mosscap,
  title={Mosscap prompt injection challenge},
  author={{Lakera AI}},
  year={2023}
}

@article{ledoit2004well,
  title={A well-conditioned estimator for large-dimensional covariance matrices},
  author={Ledoit, Olivier and Wolf, Michael},
  journal={J. Multivariate Analysis},
  volume={88},
  number={2},
  pages={365--411},
  year={2004}
}

@inproceedings{leviathan2023fast,
  title={Fast inference from transformers via speculative decoding},
  author={Leviathan, Yaniv and Kalman, Matan and Matias, Yossi},
  booktitle={ICML},
  year={2023}
}

@misc{lian2023openorca,
  title={{OpenOrca}: An open dataset of {GPT}-augmented {FLAN} reasoning traces},
  author={Lian, Wing and others},
  year={2023}
}

@article{liu2024formalizing,
  title={Formalizing and benchmarking prompt injection attacks and defenses},
  author={Liu, Yupei and Jia, Yuqi and Geng, Runpeng and Jia, Jinyuan and Gong, Neil Zhenqiang},
  journal={USENIX Security},
  year={2024}
}

@article{magar2022data,
  title={Data contamination: From memorization to exploitation},
  author={Magar, Inbal and Schwartz, Roy},
  journal={ACL},
  year={2022}
}

@article{marks2024geometry,
  title={The geometry of truth: Emergent linear structure in {LLM} representations of true/false datasets},
  author={Marks, Samuel and Tegmark, Max},
  journal={COLM},
  year={2024}
}

@article{mazeika2024harmbench,
  title={{HarmBench}: A standardized evaluation framework for automated red teaming and robust refusal},
  author={Mazeika, Mantas and others},
  journal={ICML},
  year={2024}
}

@misc{miraoui2024jailbreak,
  title={Jailbreak prompts collection},
  author={Miraoui, Yanis},
  year={2024},
  howpublished={Hugging Face}
}

@article{padhi2024granite,
  title={{Granite Guardian}},
  author={Padhi, Inkit and others},
  journal={arXiv:2412.07724},
  year={2024}
}

@article{perez2022ignore,
  title={Ignore previous prompt: Attack techniques for language models},
  author={Perez, F{\'a}bio and Ribeiro, Ian},
  journal={NeurIPS ML Safety Workshop},
  year={2022}
}

@article{radharapu2023aart,
  title={{AART}: {AI}-assisted red-teaming with diverse data generation for new {LLM}-powered applications},
  author={Radharapu, Bhaktipriya and Robinson, Kevin and Aroyo, Lora and Lahoti, Preethi},
  journal={EMNLP Industry Track},
  year={2023}
}

@misc{safeguard2024,
  title={Safeguard benign prompts corpus},
  author={{Safeguard AI}},
  year={2024}
}

@article{sainz2023nlp,
  title={{NLP} evaluation in trouble: On the need to measure {LLM} data contamination for each benchmark},
  author={Sainz, Oscar and Campos, Jon Ander and Garc{\'\i}a-Ferrero, Iker and Etxaniz, Julen and de Lacalle, Oier Lopez and Agirre, Eneko},
  journal={EMNLP Findings},
  year={2023}
}

@inproceedings{schwartz2020right,
  title={The right tool for the job: Matching model and instance complexities},
  author={Schwartz, Roy and Stanovsky, Gabriel and Swayamdipta, Swabha and Dodge, Jesse and Smith, Noah A.},
  booktitle={ACL},
  year={2020}
}

@misc{softage2024,
  title={{SoftAge} instruction-following prompts},
  author={{SoftAge AI}},
  year={2024}
}

@article{song2020mpnet,
  title={{MPNet}: Masked and permuted pre-training for language understanding},
  author={Song, Kaitao and Tan, Xu and Qin, Tao and Lu, Jianfeng and Liu, Tie-Yan},
  journal={NeurIPS},
  year={2020}
}

@inproceedings{teerapittayanon2016branchynet,
  title={{BranchyNet}: Fast inference via early exiting from deep neural networks},
  author={Teerapittayanon, Surat and McDanel, Bradley and Kung, H.T.},
  booktitle={ICPR},
  year={2016}
}

@article{wang2023donotanswer,
  title={Do-Not-Answer: A dataset for evaluating safeguards in {LLMs}},
  author={Wang, Yuxia and Li, Haonan and Han, Xudong and Nakov, Preslav and Baldwin, Timothy},
  journal={EACL Findings},
  year={2023}
}

@article{wang2024survey,
  title={A survey on large language model based autonomous agents},
  author={Wang, Lei and others},
  journal={Frontiers of Computer Science},
  year={2024}
}

@article{xi2023rise,
  title={The rise and potential of large language model based agents: A survey},
  author={Xi, Zhiheng and others},
  journal={arXiv:2309.07864},
  year={2023}
}

@inproceedings{xin2020deebert,
  title={{DeeBERT}: Dynamic early exiting for accelerating {BERT} inference},
  author={Xin, Ji and Tang, Raphael and Lee, Jaejun and Yu, Yaoliang and Lin, Jimmy},
  booktitle={ACL},
  year={2020}
}

@article{yi2024benchmarking,
  title={Benchmarking and defending against indirect prompt injection attacks on large language models},
  author={Yi, Jingwei and Xie, Yueqi and Zhu, Bin and Kiciman, Emre and Sun, Guangzhong and Xie, Xing and Wu, Fangzhao},
  journal={KDD},
  year={2024}
}

@article{zeng2024shieldgemma,
  title={{ShieldGemma}: Generative {AI} content moderation based on {Gemma}},
  author={Zeng, Wenjun and others},
  journal={arXiv:2407.21772},
  year={2024}
}

@article{zhan2024injecagent,
  title={{InjecAgent}: Benchmarking indirect prompt injections in tool-integrated {LLM} agents},
  author={Zhan, Qiusi and Liang, Zhixiang and Ying, Zifan and Kang, Daniel},
  journal={ACL Findings},
  year={2024}
}

@article{zou2023universal,
  title={Universal and transferable adversarial attacks on aligned language models},
  author={Zou, Andy and Wang, Zifan and Carlini, Nicholas and Nasr, Milad and Kolter, J. Zico and Fredrikson, Matt},
  journal={arXiv:2307.15043},
  year={2023}
}

@article{zou2023representation,
  title={Representation engineering: A top-down approach to {AI} transparency},
  author={Zou, Andy and others},
  journal={arXiv:2310.01405},
  year={2023}
}

\appendix

\section{Full Leakage Audit}
\label{app:leak}

This appendix reports the full leakage audit summarized in Section~\ref{sec:data}. The audit was applied to every candidate source in the UPIA (direct) and XPIA (indirect) pools, each treated as held-out and compared against the union of all other same-class sources. Four leakage statistics are reported: 7-gram contamination (fraction of held-out documents sharing at least one 7-gram with the train pool), 13-gram contamination (same with a stricter 13-token shingle, following large-scale pretraining contamination audits \citep{brown2020language}), and MPNet \citep{song2020mpnet} nearest-neighbour duplicate rates at cosine-similarity thresholds 0.70 and 0.85. MPNet is a sentence-embedding model that maps texts to vectors; high cosine similarity means that two prompts are semantically close. The threshold 0.85 corresponds to a strong paraphrase match. Held-out evaluation uses all prompts from the held-out source. A source is admitted to the experimental pool only if both contam\_13gram and the MPNet duplicate rate at threshold 0.85 are at most 5 percent (0.05 as a fraction). Table~\ref{tab:leak} lists every source admitted into either pool, together with the four metric values.

\begin{table*}[t]
\centering
\footnotesize
\setlength{\tabcolsep}{3pt}
\begin{tabular}{llcrrrr}
\toprule
Pool & Source & Class & 7-g \% & 13-g \% & cos$\geq$.70 \% & cos$\geq$.85 \% \\
\midrule
UPIA & aart           & att & 4.73 & 0.00 & 0.13 & 0.00 \\
UPIA & orbench\_toxic & att & 1.22 & 0.00 & 0.40 & 0.00 \\
UPIA & beavertails    & att & 2.13 & 0.00 & 0.27 & 0.00 \\
UPIA & donotanswer    & att & 1.10 & 0.00 & 0.10 & 0.00 \\
UPIA & yanismiraoui   & att & 0.10 & 0.00 & 0.19 & 0.00 \\
UPIA & mosscap        & att & 0.33 & 0.00 & 0.07 & 0.00 \\
UPIA & enron          & ben & 0.13 & 0.00 & 0.07 & 0.00 \\
UPIA & softage        & ben & 1.20 & 0.00 & 1.00 & 0.00 \\
UPIA & 10k\_prompts   & ben & 0.80 & 0.07 & 1.93 & 0.27 \\
UPIA & dolly15k       & ben & 0.47 & 0.07 & 3.33 & 0.87 \\
XPIA & XPIA           & att & 0.00 & 0.00 & 0.00 & 0.00 \\
XPIA & BIPIA          & att & 0.00 & 0.00 & 0.00 & 0.00 \\
XPIA & InjecAgent     & att & 0.00 & 0.00 & 0.00 & 0.00 \\
XPIA & AgentDojo      & att & 0.00 & 0.00 & 0.00 & 0.00 \\
XPIA & dolly15k       & ben & 0.30 & 0.00 & 0.40 & 0.00 \\
XPIA & enron          & ben & 0.20 & 0.00 & 0.00 & 0.00 \\
XPIA & softage        & ben & 1.10 & 0.00 & 0.30 & 0.00 \\
XPIA & 10k\_prompts   & ben & 0.67 & 0.00 & 0.50 & 0.00 \\
\bottomrule
\end{tabular}
\caption{Leakage audit over the 10 UPIA and 8 XPIA sources used in the experiments (four benign sources are shared between the pools). All values are percentages. Every admitted source has 13-gram contamination at most 0.07 percent and MPNet duplicate rate at threshold 0.85 at most 0.87 percent, well below the 5 percent admission ceiling on both axes. ``att'' denotes attack, ``ben'' denotes benign.}
\label{tab:leak}
\end{table*}

All sources admitted to the experimental pools clear both admission criteria with substantial margin: the worst-case 13-gram contamination among admitted sources is 0.07 percent (dolly15k, 10k\_prompts) and the worst-case MPNet duplicate rate at threshold 0.85 is 0.87 percent (dolly15k under the UPIA pool). The four XPIA attack sources (XPIA, BIPIA, InjecAgent, AgentDojo) show zero overlap on every metric, suggesting little cross-source reuse under this audit.

Several screened sources were excluded because they failed at least one threshold. On the attack side, AdvBench \citep{zou2023universal} has low 13-gram contamination but a paraphrase duplicate rate at threshold 0.85 of about 15 percent. HarmBench \citep{mazeika2024harmbench} shows similar paraphrase overlap at about 10 percent, and WildJailbreak \citep{jiang2024wildjailbreak} is borderline at about 3.5 percent. On the benign side, OpenOrca \citep{lian2023openorca}, the Deepset benign split \citep{deepset2023}, and Safeguard \citep{safeguard2024} fail the lexical-overlap threshold because of templated instruction-tuning text. Per-source rejection statistics are recorded in the project artifact and are omitted here for space.

\section{Tolerance Ablation for the Pareto Layer}
\label{app:eps}

The headline results use the Pareto layer with tolerance $\varepsilon = 0.005$ (Section~\ref{sec:layer}). This appendix reports outer-LOSO results for every (host, pool) cell at $\varepsilon \in \{0, 0.005, 0.010, 0.020\}$. By construction, the selected layer's aggregated inner-LOSO BAcc is at most $\varepsilon$ below the audit-best layer. This bound applies to training-pool BAcc used for layer selection, not to held-out outer-LOSO BAcc. The question of interest is how the chosen layer and outer-LOSO BAcc change as $\varepsilon$ grows.

\begin{table*}[!t]
\centering
\footnotesize
\setlength{\tabcolsep}{4pt}
\begin{tabular}{llrrrrrrrr}
\toprule
Host & Pool & $L(0)$ & BAcc(0) & $L(.005)$ & BAcc(.005) & $L(.010)$ & BAcc(.010) & $L(.020)$ & BAcc(.020) \\
\midrule
LlamaGuard-3        & UPIA & 16 & 0.7958 & 16 & 0.7958 & 16 & 0.7958 & 16 & 0.7958 \\
LlamaGuard-3        & XPIA & 24 & 0.9114 & 20 & 0.9124 & 20 & 0.9124 & 16 & 0.9032 \\
Granite-Guardian-8B & UPIA & 30 & 0.8379 & 30 & 0.8379 & 30 & 0.8379 & 25 & 0.8270 \\
Granite-Guardian-8B & XPIA & 20 & 0.9237 & 20 & 0.9237 & 20 & 0.9237 & 15 & 0.9126 \\
ShieldGemma-9B      & UPIA & 30 & 0.8421 & 24 & 0.8426 & 24 & 0.8426 & 24 & 0.8426 \\
ShieldGemma-9B      & XPIA & 36 & 0.9096 & 24 & 0.9077 & 24 & 0.9077 & 24 & 0.9077 \\
WildGuard-7B        & UPIA & 28 & 0.7820 & 24 & 0.7737 & 16 & 0.7692 & 16 & 0.7692 \\
WildGuard-7B        & XPIA & 16 & 0.9068 & 16 & 0.9068 & 16 & 0.9068 & 16 & 0.9068 \\
\bottomrule
\end{tabular}
\caption{Tolerance ablation for the Pareto layer policy. For each (host, pool) cell, the table reports the selected layer $L(\varepsilon)$ and outer-LOSO balanced accuracy at $\varepsilon \in \{0, 0.005, 0.010, 0.020\}$. Cells where $\varepsilon = 0.005$ (the headline configuration) selects a strictly shallower layer than $\varepsilon = 0$ identify where shallower deployment layers contribute to the latency reduction reported in Section~\ref{sec:exp}.}
\label{tab:eps}
\end{table*}

The largest movements happen at $\varepsilon = 0.020$, where 4 cells shift to a strictly shallower layer; the worst outer BAcc cost is 0.011 (LlamaGuard-3 / XPIA, 0.9124 $\rightarrow$ 0.9032). At $\varepsilon = 0.005$ and $\varepsilon = 0.010$, the same 4 cells shift shallower: LlamaGuard-3 / XPIA, ShieldGemma-9B / UPIA, ShieldGemma-9B / XPIA, and WildGuard-7B / UPIA. The maximum outer-BAcc drop across all 8 cells at $\varepsilon = 0.005$ is 0.83 percentage points (WildGuard-7B / UPIA), in line with the inner-BAcc tolerance. The headline configuration $\varepsilon = 0.005$ therefore captures much of the available depth reduction at negligible quality cost.

\end{document}